\documentclass[a4paper,11pt]{article}
\pdfoutput=1 

\usepackage{jcappub} 

\usepackage[T1]{fontenc} 

\usepackage{graphicx}
\usepackage{multirow}
\usepackage{booktabs}
\usepackage[table]{xcolor}

\hypersetup{
  colorlinks=true,
  linkcolor=red,
  urlcolor=blue,
  citecolor=red, 
  pdfauthor={Name}
}

\title{Impact of QCD uncertainties on antiproton spectra from dark-matter annihilation}

\author[a]{Adil Jueid,}
\author[b]{Jochem Kip,}
\author[c]{Roberto Ruiz de Austri}
\author[d]{and Peter Skands}

\affiliation[a]{Center for Theoretical Physics of the Universe, Institute for Basic Science (IBS),\\ Daejeon, 34126, Republic of Korea}
\affiliation[b]{Institute for Mathematics, Astrophysics and Particle Physics, Radboud University, \\
Nijmegen, Heyendaalseweg 135, Nijmegen, the Netherlands}
\affiliation[c]{Instituto de F\'{\i}sica Corpuscular, CSIC-Universitat de Val\`encia, \\ E-46980 Paterna, Valencia, Spain}
\affiliation[d]{School of Physics and Astronomy, Monash University, \\ Wellington Rd,  Clayton VIC-3800, Australia}

\emailAdd{adiljueid@ibs.re.kr} 
\emailAdd{jochem.kip@ru.nl}
\emailAdd{rruiz@ific.uv.es}
\emailAdd{peter.skands@monash.edu}

\abstract{Dark-matter particles that annihilate or decay can undergo complex sequences of processes, including strong and electromagnetic radiation, hadronisation, and hadron decays, before particles that are stable on astrophysical time scales are produced. Antiprotons produced in this way may leave footprints in experiments such as AMS--02. Several groups have reported an excess of events in the antiproton flux in the rigidity range of $10$--$20$ GV. However, the theoretical modeling of baryon production is not straightforward and relies in part on phenomenological models in Monte Carlo event generators. 
In this work, we assess the impact of QCD uncertainties on the spectra of antiprotons from dark-matter annihilation. As a proof-of-principle, we show that for a two-parameter model that depends only on the thermally-averaged annihilation cross section ($\langle \sigma v \rangle$) and the dark-matter mass ($M_\chi$), QCD uncertainties can affect the best-fit mass 
by up to $\sim 14 \%$ (with large uncertainties for large DM masses), depending on the choice of $M_\chi$ and the annihilation channel ($b\bar{b}$ or $W^+ W^-$), 
and $\langle \sigma v \rangle$ by up to $\sim 10\%$. For comparison, changes to the underlying diffusion parameters are found to be within $1\%$--$5\%$, and the results are also quite resilient to the choice of  cosmic-ray propagation model. These findings indicate that QCD uncertainties need to be included in future DM analyses. To facilitate full-fledged analyses, we provide the spectra in tabulated form including QCD uncertainties and code snippets to perform mass interpolations and quick DM fits. The code can be found in this \href{https://github.com/ajueid/qcd-dm.github.io.git}{github} repository.}


\begin{document}
\maketitle
\flushbottom

\section{Introduction}
\label{sec:intro}

In the Cosmological Standard Model, dark matter (DM) constitutes about $85\%$ of the matter in the universe (see {\it e.g.} \cite{Bertone:2004pz}). Analyses of the cosmic microwave background prefer DM particles that are nonrelativistic in the era of galaxy formation. This requirement is relatively straightforward to realize in particle-physics models by extending the Standard Model (SM) with weakly interacting massive particles (WIMPs). Such scenarios can reach excellent agreement with the measured relic density, $\Omega_{\rm DM} h^2 = 0.1188\pm 0.0010$~\cite{Ade:2015xua}, and typically predict that WIMPs can annihilate (or decay) into SM states, which --- via often complex sequences of decays, radiation, and hadronisation --- produce observable final-state particles such as photons, anti-protons, antineutrinos, or positrons. 
Contributions of such DM annihilation products to cosmic-ray (CR) fluxes may be observable in experiments like the Fermi Large Area Telescope (\textsc{Fermi--LAT}), or the Alpha Magnetic Spectrometer~(AMS--02). \\
After the discovery of secondary CR antiprotons in 1979, an excess over the astrophysical background was observed in the antiproton-to-proton ratio \cite{Golden:1979bw, Buffington:1981zz} which at the time was explained by a photino DM with mass of a few GeV \cite{Silk:1984zy}. Despite the unprecedented precision on the measurement of the antiproton flux performed by the AMS--02 collaboration \cite{AMS:2016oqu}, the excess seems to be still alive especially in the rigidity range $10$--$20$ GV (see {\it e.g.} \cite{Cuoco:2016eej,Cui:2016ppb,Cuoco:2017rxb,Reinert:2017aga}). Interestingly, DM interpretations of this excess can be consistent with those of the so-called Galactic Center Excess (GCE) in gamma rays \cite{Goodenough:2009gk, Vitale:2009hr, Hooper:2010mq, Gordon:2013vta, Hooper:2011ti, Daylan:2014rsa, Calore:2014xka, Caron:2015wda, vanBeekveld:2016hbo, achterberg2018implications}. In other words, a DM of mass around $50$--$200$~GeV and thermal annihilation cross section of about $\sim 10^{-26}~{\rm cm}^2~{\rm s}^{-1}$ can address both of these anomalies. The statistical uncertainties on the antiproton flux are now considered very subleading, making the treatment of systematic uncertainties and their correlations crucial to further progress in DM analyses \cite{diMauro:2014zea,Kappl:2014hha,Kachelriess:2015wpa,Winkler:2017xor,Korsmeier:2018gcy,Heisig:2020nse,Calore:2022stf}. It is in this context that we note that, for DM masses above a few GeV, antiprotons are produced through Quantum Chromodynamics (QCD) jet fragmentation. Since the problem of confinement remains fundamentally unsolved, it is addressed either via phenomenological fits such as Fragmentation Functions (FFs), and/or explicit dynamical models embodied by Monte Carlo (MC) event generators. Both approaches involve free parameters (see, {\it e.g.}, \cite{Buckley:2011ms,Skands:2014pea,ParticleDataGroup:2020ssz}).
 The associated uncertainties on the particle fluxes in dark-matter annihilation have, however, so far been neglected in the literature. As experimental systematics at the few-percent level have caused important shifts in the antiproton excess, it is relevant to examine whether QCD uncertainties could be of a comparable magnitude to this. 
 
In this article, we perform a dedicated analysis with state-of-the-art MC tools which shows that QCD uncertainties can alter, and even dominate, the particle-physics component of the overall error budget for antiproton DM fits. This is quite different to the case of gamma rays for which QCD uncertainties are below $10\%$ in the peak region~\cite{Amoroso:2018qga} (see also \cite{Amoroso:2020mjm,Jueid:2021dlz} for short summaries). 
We expect this to have a significant impact on DM  interpretations of the AMS--02 excess. We first perform several fits of the fragmentation-function parameters in \textsc{Pythia} 8.244 \cite{Sjostrand:2014zea} to salient LEP measurements, starting from the default  ``\textsc{Monash}'' tune \cite{Skands:2014pea}. We then discuss the different sources of QCD uncertainties on the antiproton production from DM annihilation and show that, in our analysis they are of similar size as the experimental errors on the AMS--02 data, especially in the low rigidity range. As a first step towards quantifying the impact on DM fits, we study the effects on the best-fit points of both the mass of the DM as well as the thermally-averaged annihilation cross section for a few annihilation channels and for fitted diffusion parameters of the CR propagation model defined in DRAGON2. We encourage complementary measurements to be developed, e.g., at the Large Hadron Collider (LHC), which could be used to further improve the quality of the theoretical models for baryon production.  The rest of this article is organised as follows. In section \ref{sec:antiprotons}, we discuss the modeling of antiproton spectra in \textsc{Pythia}~8 and briefly highlight differences in the theory predictions of the state-of-art multi-purpose MC event generators: \textsc{Herwig}~7, \textsc{Pythia}~8 and \textsc{Sherpa}~2. In section \ref{sec:QCDuncertainties}, we present the results of a new tune of the Lund fragmentation-function parameters in \textsc{Pythia}~8 and estimate the QCD uncertainties for a few DM masses. In section \ref{sec:DM_unc}, we discuss the impact on DM observables. Finally we conclude in section \ref{sec:conclusion}. 



\section{Antiprotons from DM annihilation}
\label{sec:antiprotons}

For a generic production of a final state $f$\footnote{$f$ can be any parton-level final state, typically pairs such as $f \in [\gamma\gamma,\ell\ell, q\bar{q}, t\bar{t}, gg, WW, ZZ, hh, \ldots]$. Since this work focuses on antiprotons, we disregard leptonic final states and QED-like processes connected to them.} in DM annihilation or decay, the general expression of the antiproton flux at the position of Earth can be written as \cite{Cirelli:2010xx}

\begin{eqnarray}
\frac{{\rm d}\Phi_{\bar{p}}}{{\rm d}E_{\rm kin}}(E_{\rm kin}, r_{\odot}) \equiv \frac{v_{\bar{p}}}{4\pi} \bigg(\frac{\rho_\odot}{M_\chi}\bigg)^k \mathcal{R}(E_{\rm kin}) \sum_f k^{-1} \mathcal{P}_{i\to f}(g_\chi; M_\chi) \bigg(\frac{{\rm d}N}{{\rm d}E_{\rm kin}}\bigg)_{f\to\bar{p}},
\label{eq:pbar:flux}
\end{eqnarray}
where $k=2~(1)$ for annihilation~(decay) process, $\rho_{\odot}$ sets the normalisation of the DM density in terms of its reference value at the Sun, $E_{\rm kin}$ is the kinetic energy of the antiproton and $v_{\bar{p}}$ is its velocity. $\mathcal{R}(E_{\rm kin})$ encodes all the astrophysical factors on both the production and the propagation of antiprotons including the shape of the DM distribution.
Particle physics enters in the last two terms in the right hand side of equation \eqref{eq:pbar:flux}: {\it (i)} The model-dependent term is encoded in $\mathcal{P}_{i \to f}(g_\chi; M_\chi)$ that gives the transition rate for DM into the final state $f$ which can be either the thermal annihilation cross section today ($\langle \sigma v \rangle_{\chi \chi \to f}$) for the annihilation or the partial decay width ($\Gamma_{\chi \to f}$) for the decay of DM and {\it (ii)} The model-independent part which represents the differential distribution of the antiproton on the kinetic energy. The model-independent contribution to the antiproton flux is usually computed using MC event generators. Coloured particles produced in DM processes will undergo QCD bremsstrahlung wherein additional coloured particles are produced and whose multiplicity depend on how far the parent particles are from their production threshold. The rate of QCD processes is controlled by the effective value of the strong coupling constant evaluated a scale proportional to the shower evolution variable \footnote{We note that the value of $\alpha_S(M_Z)$ in \textsc{Pythia}~8 is larger than $\alpha_S(M_Z)^{\overline{{\rm MS}}}$ by about $20\%$ \cite{Skands:2010ak,Skands:2014pea}.}. At a scale $Q_{\rm IR} \simeq \mathcal{O}(1)~{\rm GeV} \ll Q_{\rm UV}$, any coloured particle must hadronise to produce a set of colourless hadrons. This process, called fragmentation is modeled within \textsc{Pythia}~8 with the Lund string model \cite{Andersson:1983ia, Sjostrand:1982fn, Sjostrand:1984ic} whose longitudinal properties (in lightcone coordinates) are encoded in the following fragmentation function: 
\begin{eqnarray}
f_\mathrm{Lund}(z,m_{\perp h}) & \propto & \frac{(1-z)^a}{z}\exp\left(\frac{-b m_{\perp h}^2}{z}\right)~,
\label{eq:fz}
\end{eqnarray}
with $m_{\perp h} \equiv \sqrt{m_h^2 + p_{\perp h}^2}$ being the ``transverse mass'' of the hadron $h$ with mass $m_h$ and transverse momentum  $p_{\perp h}$ relative to the string axis. In the standard Lund model, the latter is assumed to follow a 2D Gaussian distribution. For each iterative string fragmentation process, $f_\mathrm{Lund}(z)$ gives the probability that a hadron $h$ with $m_{\perp h}$ and energy fraction $z \in [0, 1]$ is produced. For mesons, it depends on three tunable parameters: the longitudinal $a$ and $b$ parameters in equation \eqref{eq:fz}, and $\sigma_\perp \approx \sqrt{\langle p_{\perp h}^2 \rangle}/\sqrt{2}$ which controls the transverse-momentum distribution. For baryons, a fourth parameter enters, with $a \to a + a_{\rm QQ}$. The $a$ and $b$ parameters act to suppress the relative rates of very hard and very soft hadrons, respectively, and in practice are highly correlated~\cite{Amoroso:2018qga}. For this reason a new parameterization of $f_\mathrm{Lund}(z)$ was introduced wherein $b$ is replaced by 
$\langle z_\rho \rangle$, the average longitudinal momentum fraction taken by a $\rho$ meson; $\langle z_\rho \rangle = \int_0^1 {\rm d}z \ z f(z, \langle m_{\perp \rho} \rangle)$. 

\begin{figure}[!t]
\centering
\vspace{-1cm}
\includegraphics[width=0.95\linewidth]{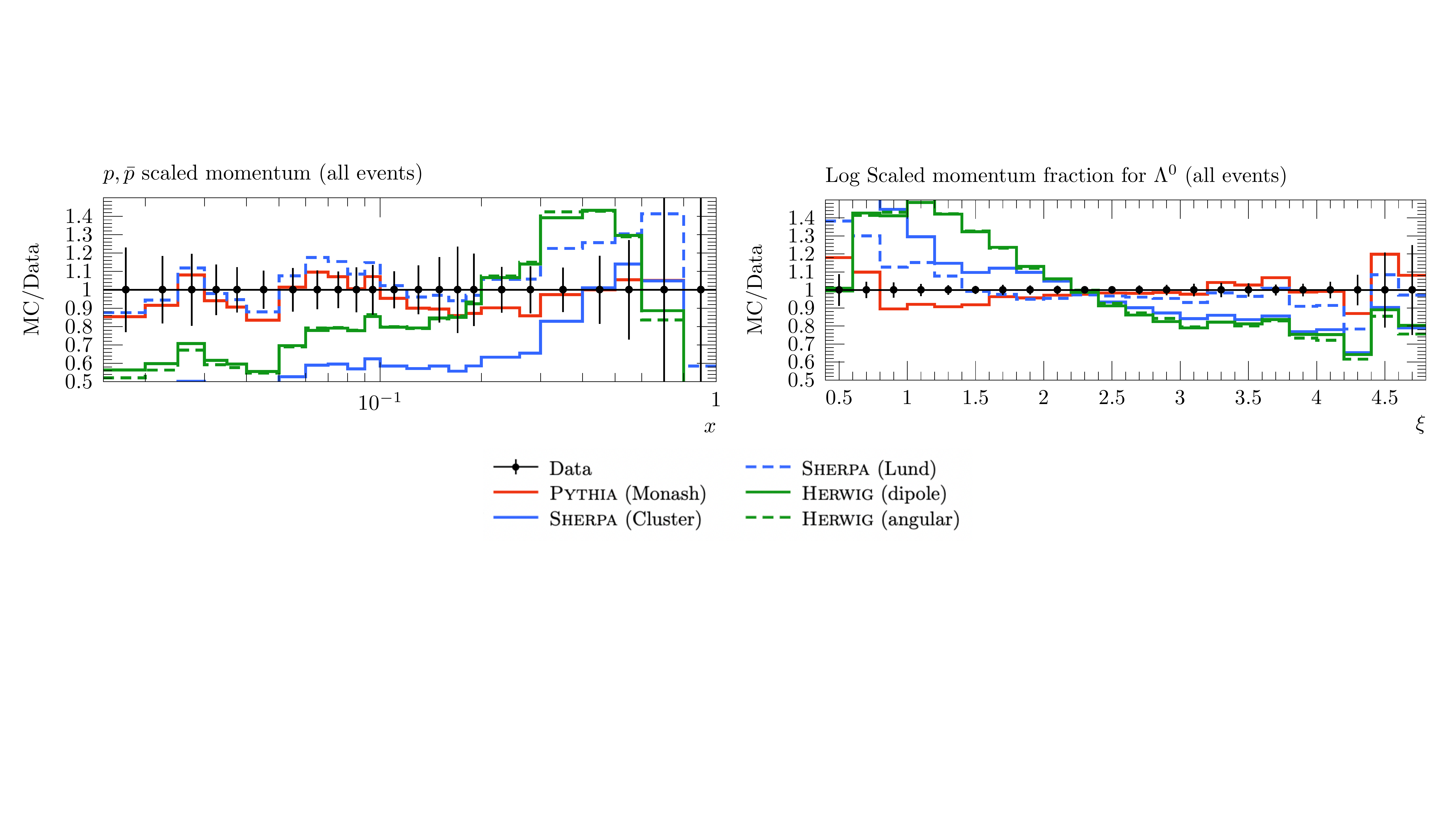}
\vspace{-2.5cm}
\caption{Comparison between the theory predictions and the experimental measurements of $p/\bar{p}$ scaled-momentum (left) and $\Lambda^0$ Log of scaled momentum at LEP. \textsc{Pythia}~8 with the default Monash tune is shown in red, \textsc{Sherpa}~2  is shown in blue for the cluster (solid) and Lund (dashed) hadronisation models while the result of \textsc{Herwig}~7 is shown in green for dipole (solid) and angular-based (dashed) parton-shower algorithms. Data is taken from \textsc{Opal} \cite{Abreu:1998vq} (left) and \textsc{Aleph} \cite{Barate:1999gb} (right).}
\label{fig:comparison:MC}
\end{figure}

The fragmentation-function parameters cannot be computed from first principles. But exploiting the fact that hadronisation occurs at very long distance scales as compared to high-energy scattering processes, the factorization theorem holds, and measurements of hadron spectra at {\it e.g.} the Large Electron Positron Collider (LEP) can be used to constrain them. The same principle implies that those parameters can then also be used to make predictions for hadronic DM annihilation processes (at least for DM masses above a few GeV); this is sometimes referred to as jet universality. At LEP, hadronic $Z$ decays produced (anti-)protons either directly from the fragmentation of quarks/gluons (dubbed primary) or from decays of heavier baryons (dubbed secondary). Secondary (anti)protons come mainly from decays of five baryons: $\Lambda^0, \Delta^{++}, \Delta^{+}, \Delta^0$ and $\Sigma^{\pm}$, with the following fractions to the total number of produced antiprotons: $R_{\Lambda^0 \to p} \approx 22\%, R_{\Delta^{++} \to p} \approx 10\%, R_{\Delta^+ \to p} \approx 8\%, R_{\Delta^0 \to p} \approx 3\%$ and $R_{\Sigma^{\pm} \to p} \approx 5\%$\footnote{This finding also holds true for antiprotons from dark matter annihilation except that there is a significant contribution from antineutron decays, which are stable at LEP time scales while they are not in dark-matter indirect detection experiments. Nevertheless, if one considers the antiprotons that are not produced from antineutron decays (roughly $50\%$ of the total), the contribution from the other sources is almost independent of the dark matter mass and annihilation final state.}. We close this section by discussing the agreement between the theory predictions of three state-of-art MC event generators and the experimental measurements of the baryon spectra at LEP (proton and $\Lambda^0$). In figure \ref{fig:comparison:MC}, we show the comparison between the predictions of ({\it i}) \textsc{Pythia}~8 version 307 \cite{Bierlich:2022pfr} with the default \textsc{Monash} tune \cite{Skands:2014pea}, ({\it ii}) \textsc{Sherpa} version 2.2.12 \cite{Gleisberg:2008ta} with the CSS shower model \cite{Schumann:2007mg} using the cluster hadronisation model \cite{Winter:2003tt} and the Lund string model based on \textsc{Pythia}~6 \cite{Sjostrand:2006za} and ({\it iii}) \textsc{Herwig} version 7.2.3 \cite{Bellm:2015jjp} with two radiation models: the angular \cite{Gieseke:2003rz} and the dipole-type \cite{Platzer:2009jq, Platzer:2011bc} algorithms for two key observables used in our fits. We can see that the predictions of \textsc{Herwig}~7 and \textsc{Sherpa} based on the cluster models disagree with both the \textsc{Pythia}~8 prediction as well as with the experimental measurements. Since the disagreement can reach up to $50\%$, we conclude that the envelope spanned by the differenc MC event generators cannot define a faithful estimate of the QCD uncertainties on the baryon spectra.

\begin{table}[!t]
\setlength\tabcolsep{3pt}
  \begin{center}
    \begin{tabular}{llclcccc}
      \toprule
      \toprule
      Parameter & \textsc{Pythia8} setting & \textsc{Monash} & This work & $a_{\rm q}$ & $\langle z_\rho \rangle$ & $\sigma_{\perp}$ [GeV] & $a_{QQ}$\\
      \midrule
      \midrule
      $a_{\rm q}$ & \verb|StringZ:aLund| &  $0.68$ & $0.601_{-0.038}^{+0.038}$ & \cellcolor{blue!30}$1.000$\\
      $\langle z_\rho \rangle$ & \verb|StringZ:avgZLund| & $(0.55)$ & $0.540_{-0.004}^{+0.004}$ & \cellcolor{red!30}$0.718$ & \cellcolor{blue!30}$1.000$ & \\
      $\sigma_{\perp}$~[GeV] & \verb|StringPT:Sigma|  & $0.335$ & $0.307_{-0.002}^{+0.002}$ & \cellcolor{red!10}$0.057$ & \cellcolor{cyan!30}$-0.270$ & \cellcolor{blue!30}$1.000$ &  \\
      $a_{QQ}$ & \verb|StringZ:aExtraDiquark| & $0.97$ & $1.671_{-0.196}^{+0.196}$ & \cellcolor{red!20}$0.415$ & \cellcolor{red!35}$0.816$ & \cellcolor{cyan!25}$-0.204$ & \cellcolor{blue!30}$1.000$ \\ 
      $b$ & \verb|StringZ:bLund| & $0.98$ & $(0.897)$ \\
      \bottomrule
      \bottomrule
    \end{tabular}
  \end{center}
  \caption{\label{tab:tunes:results} The fit results for the parameters of the fragmentation function in \textsc{Pythia}~8. The errors on the parameters correspond to the \textsc{Migrad} errors estimated at the fit. For comparison we show the corresponding values of the parameters in the baseline \textsc{Monash} tune. The last four columns show the symmetric correlation matrix. The goodness-of-fit per degrees of freedom ($\chi^2/N_{\rm df}$) is $\chi^2/N_{\rm df} = 676.69/852$ where $N_{\rm df} = N_{\rm bins} - N_{\rm params}$.}
\end{table}

\section{QCD uncertainties on antiproton fluxes}
\label{sec:QCDuncertainties}

\begin{figure}[!h]
    \centering
    \includegraphics[width=0.6\linewidth]{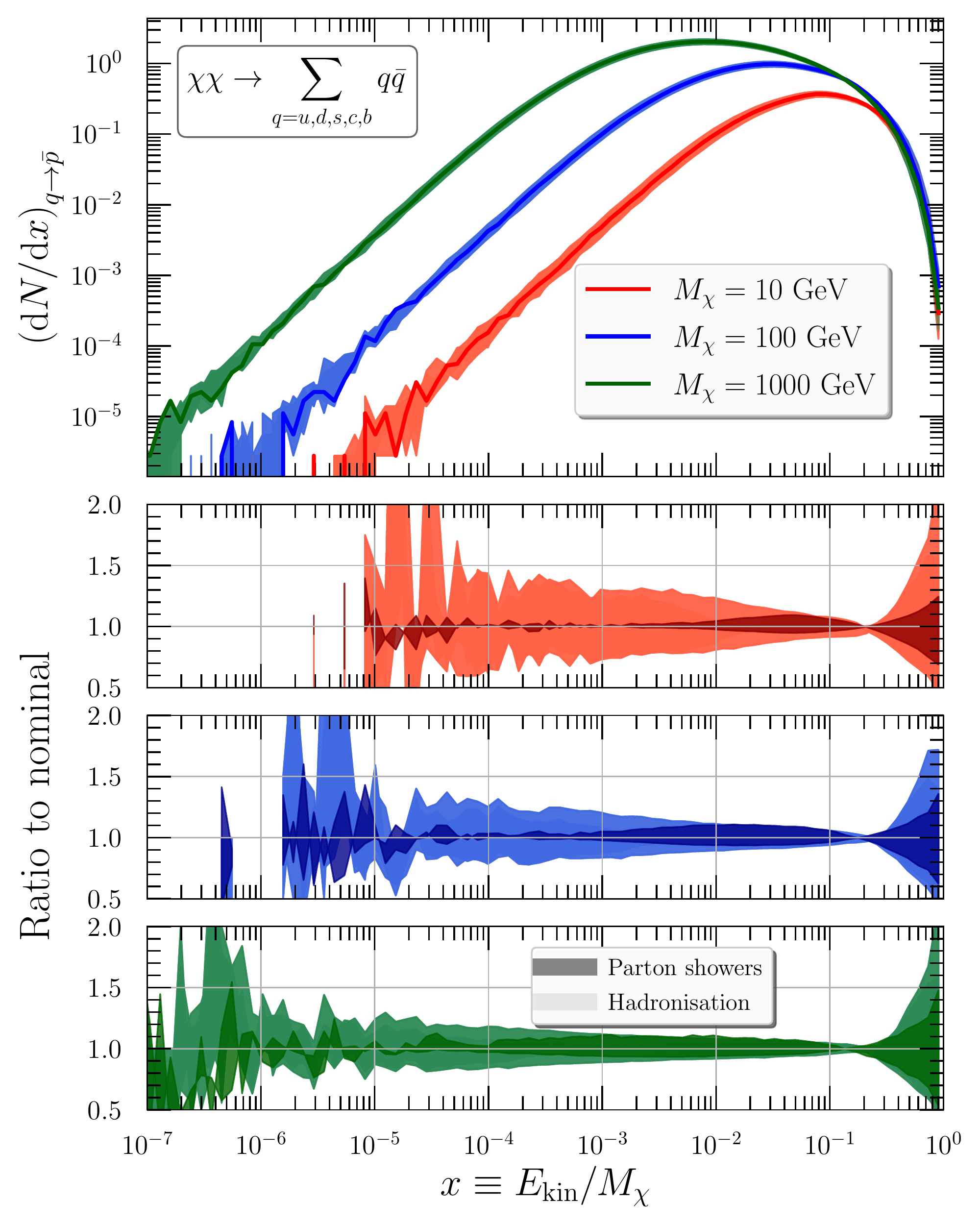}
    \caption{Antiproton differential distribution in $x \equiv E_{\rm kin}/M_\chi$ for dark matter annihilation into $q\bar{q}$ for $M_{\chi} = 10~{\rm GeV}$ (red), $M_\chi = 100~{\rm GeV}$ (blue) and $M_\chi = 1000~{\rm GeV}$ (dark green). The three lower panels show the ratio to the nominal prediction for each DM mass. The nominal prediction for each case were estimated using the results in Table \ref{tab:tunes:results}. The light (dark) error bands correspond to uncertainties on the predictions from hadronisation (parton showers). The hadronisation uncertainties correspond to the $2\sigma$ eigentunes (see the text). Here, we assume that the dark matter is annihilating to all the SM quarks but the top quark with equal probabilities, {\it i.e.} ${\rm BR}(\chi\chi \to q\bar{q}) \equiv \frac{\sigma(\chi\chi \to q\bar{q})}{\sum_q \sigma(\chi\chi \to q\bar{q}} =  20\%$ for $q = u,d,s,c,b$.}
    \label{fig:antip:spectra}
\end{figure}

To assess properly the QCD uncertainties on the antiproton spectra, we  first perform several retunings of \textsc{Pythia}~8 using a set of constraining data from LEP. A complete global analysis would involve measurements not only of proton/antiproton spectra but also those of hyperons and $\Delta$ baryons which decay to protons. There are, however, no measurements of $\Delta$ or $\Sigma$ baryon spectra with any significant constraining power, so for antiproton production via decays we are forced to focus solely on the $\Lambda$ component. To ensure that the fits remain consistent with overall event properties as well, we also include spectra of light mesons, mean charged-particle multiplicities and multiplicity distributions, and the thrust and $C$-parameter event shapes, using all relevant measurements available in \textsc{Rivet}~3.1.3 \cite{Bierlich:2019rhm}, specifically those by  \textsc{Aleph} \cite{Buskulic:1994ft,Barate:1996fi, Barate:1999gb,Heister:2003aj}, \textsc{Delphi} \cite{Abreu:1992gp,Abreu:1993mm, Adam:1995rf,Abreu:1995cu, Abreu:1998vq}, \textsc{L3} \cite{Achard:2004sv} and \textsc{Opal} \cite{Acton:1991aa,Akers:1994ez, Alexander:1996qj,Ackerstaff:1998hz,Abbiendi:2004qz}. The fit is performed using \textsc{Professor}~2.3.3 \cite{Buckley:2009bj}, with goodness-of-fit measure
\begin{equation}
 \chi^2 = \sum_{\mathcal{O}} \sum_{b\in \mathcal{O}} \bigg(\frac{{\rm MC}_{(b)}(\{p_i\}) - {\rm Data}_{(b)}}{\Delta_b}\bigg)^2,
\label{eq:GoF}
\end{equation}
where ${\rm MC}_{(b)}(\{p_i\})$ denotes the MC prediction for observable $\mathcal{O}$ at a bin $b$, which depends on the parameter set $\{p_i\} = \{a_{\rm q}, \langle z_\rho \rangle, \sigma_\perp, a_{QQ}\}$, cast as a fourth-order polynomial interpolation (see {\it e.g.} \cite{Amoroso:2018qga} for details), and $\Delta_b$ represents the total error which is estimated as the quadratic sum of the MC statistical errors, experimental errors and a $5\%$ flat theory uncertainty; $\Delta_b = \sqrt{\Delta_{\rm MC,b}^2 + \Delta_{\rm Data,b}^2 + (0.05 \times {\rm MC}_{(b)})^2}$. The results of the tunes along with the corresponding correlation matrix are shown in Table \ref{tab:tunes:results}. We have verified that the predictions of \textsc{Pythia}~8 at the best-fit point agree very well with both the results from the baseline \textsc{Monash} tune and  with the bulk of the experimental data. 

The QCD uncertainties on the antiproton spectra can be categorised into: {\it (i)} perturbative uncertainties arising from parton-shower variations and {\it (ii)} hadronisation uncertainties on the parameters of the fragmentation function obtained at the best-fit point. The shower uncertainties are estimated using the automated method developed in \cite{Mrenna:2016sih} and include a set of next-to-leading order correction terms which act to reduce the renormalization-scale variations. For the hadronisation uncertainties, we use the \textsc{Professor} toolkit to estimate the Hessian errors which can be computed from the diagonalisation of the $\chi^2$ covariance matrix near the minimum (details can be found in {\it e.g.} \cite{Pumplin:2001ct}). One gets $2N_{\rm param} = 8$  variations, known as eigentunes, and one can associate $1\sigma$, $2\sigma$ and $3\sigma$ eigentunes to variations satisfying $\Delta \chi^2 \equiv \chi^2 - \chi^2_{\rm min} \simeq N_{\rm df}$, $4N_{\rm df}$, and $9N_{\rm df}$ respectively. Given that there are a {\it priori} unknown parametric uncertainties on the theory side and that a ``good'' uncertainty estimate should at least roughly cover the range of the experimental uncertainties, we find that the $2\sigma$ eigentunes are well suited to obtain  reasonably conservative estimates, hence we will use those as our baseline in the rest of this article. To illustrate the effects of the QCD uncertainties, we calculate the antiproton spectra for DM annihilation into $q\bar{q}$ choosing three DM masses of $10, 100$ and $1000$ GeV. The results are shown in figure \ref{fig:antip:spectra} where we can see that hadronisation uncertainties can go from $10$--$25\%$ in the peak regions to about $50$--$200\%$ in the high $x$-regions. The shower uncertainties compete with hadronisation errors in the peak and for $M_\chi \geqslant 100~{\rm GeV}$ while they are smaller for $x \geqslant 0.3$.  

\section{Effects on dark matter observables}
\label{sec:DM_unc}
In this section we quantify to what extent the QCD uncertainties affect two DM observables: the velocity-weighted cross section $\langle \sigma v \rangle$ and the DM mass $M_\chi$. A total shift in the height of the spectrum, such that it is still contained within the QCD uncertainties of the spectrum, corresponds to an uncertainty on $\langle \sigma v \rangle$. The uncertainty on the DM mass, $\Delta M_\chi$, for a DM particle of mass $M_\chi$ can be determined by finding the masses of DM particles whose spectra including QCD uncertainties match the spectrum of the DM particle with mass $M_\chi$. For both DM uncertainties we determine the upper and lower bounds by demanding that $\chi^2 / N_\text{df} \approx 1$ holds.  Thus the spectra of the parameters corresponding to the upper and lower bounds of the DM uncertainties can give the same spectra as the original $\langle \sigma v\rangle$ or $M_\chi$ when the QCD uncertainties are accounted for. We note that the uncertainty in $\langle \sigma v \rangle$ is equivalent to an uncertainty in the DM density, $\rho_\odot$. The exact conversion between $\langle \sigma v \rangle$ and $\rho_\odot$ however depends on the specific DM halo profile, thus we shall only use $\langle \sigma v \rangle$ in the following. \\
\begin{figure}[!t]
    \centering
    \includegraphics[width=\linewidth]{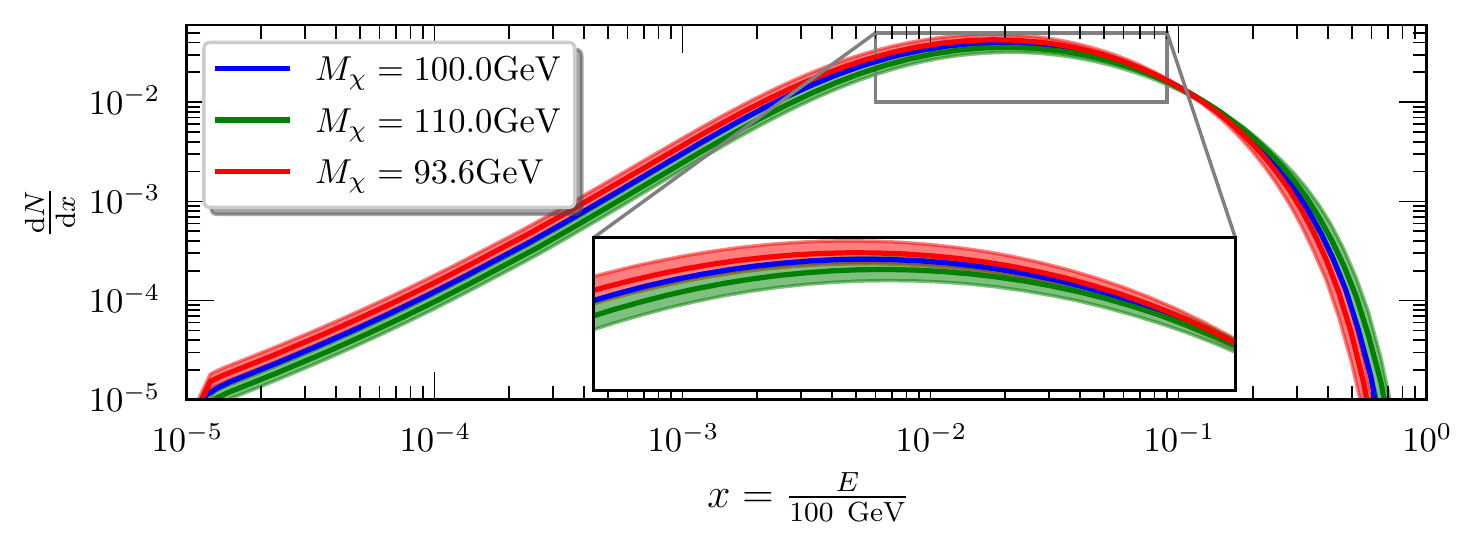}
    \caption{The nominal value of two DM particles with a mass of 100 GeV annihilating into antiprotons via a $W^+ W^-$ pair is shown in blue. The upper and lower limit of the DM masses whose antiproton spectra, again via $W^+ W^-$, encompass the nominal 100 GeV spectra in their uncertainties is shown in green and orange respectively. The peak of the spectrum is magnified for visibility purposes. Notably while the spectrum down to 1 MeV is shown, the fit only considers energies higher than 1 GeV}
    \label{fig:spectrum_uncertainty}
\end{figure}
In order to estimate the aforementioned uncertainties we have written publicly-available code, obtainable at~\href{https://github.com/ajueid/qcd-dm.github.io.git}{GitHub}, that can estimate the DM uncertainties for a wide variety of DM masses and channels. The DM antiproton spectrum after Galactic propagation is computed by diffusing the DM annihilation spectrum at each bin with pre-computed diffusion values. Thus by diffusing all bins the entire propagated spectrum is obtained. We determine the diffusion values for a specific energy bin with DRAGON 2~\cite{Evoli:2017, Evoli:2018} by using a peaked DM annihilation spectrum at the energy of the bin. The settings for DRAGON 2 are determined by fitting the proton ($p$), antiproton-over-proton ($\bar{p}/p$), and boron-over-carbon ratios ($B/C$) from AMS--02~\cite{AMS:2021} with an artificial bee colony~\cite{Dervis:2005, Akay:2012}\footnote{The propagation parameters can be found in the README of the code}. In order to make the code flexible, the pre-computed diffusion values can be supplied by the user, such that it can be tailored for specific use. Additionally, the energy range which is considered by the fit can be supplied, such that only the region of interest is fitted. We stress here explicitly that our code should not be used in place of an actual CR propagation code and should only be used in order to obtain the uncertainties on the aforementioned DM observables. Notably, our fit of the AMS-02 spectra resulted in a DM mass of $\mathcal{O}(100)~ \text{GeV}$ and additionally is only one parameter set. In order to assess the impact of different propagation parameters on the DM uncertainties, we have performed cross checks with three different propagation parameter sets: the aforementioned best-fit scenario, the DRAGON 2 default settings, and one taken from~\cite{CR_parameters:Di_Bernardo_2010}. We found that the impact of different propagation parameters results in a variation on the uncertainty of the DM observables of approximately 1-5\%. It should be noted that only a low-energy correction factor on the diffusion coefficient is implemented, so solar modulation is not fully taken into account for the pre-computed diffusion values. In the following all uncertainties are fitted by disregarding spectra energies lower than 1 GeV. While solar modulation does play a role at 1 GeV, the spectrum uncertainties are fairly uniform in this region, thus the implementation of solar modulation is not expected to change the results by a significant amount.\\
In figure \ref{fig:spectrum_uncertainty} we showcase the diffused spectrum of a 100 GeV DM particle annihilating into antiprotons via a $W^+ W^-$ pair, and two spectra which contain the 100 GeV spectrum in their uncertainty bounds, i.e. $\Delta M_\chi$. One of the main constraints determining the DM mass uncertainty is the relatively small uncertainty band at $x\approx0.2$, which can most clearly be seen in figure \ref{fig:antip:spectra}. While the diffusion of the spectra alters the pre-diffused spectrum, the small uncertainty band remains a clear feature. This region is especially important for the uncertainty on $\langle \sigma v \rangle$. \\
\begin{table}[!t]
    \centering
    \begin{tabular}{llccc}
    \toprule
    \toprule
    $M$ & Channel       & $\langle\sigma v\rangle$ [\%] & $\Delta M_\chi$ [GeV]\\
    \midrule
    \midrule
    100 & $b\bar{b}$    & $\phantom{}^{+9.2}_{-7.0}$  & $\phantom{}^{+12.8}_{-4.5}$\\
    1000 & $b\bar{b}$   & $\phantom{}^{+7.9}_{-6.3}$  & $\phantom{}^{+143.3}_{-65.5}$\\
    100 & $W^+W^-$      & $\phantom{}^{+6.7}_{-9.4}$  & $\phantom{}^{+10.0}_{-6.4}$\\
    1000 & $W^+W^-$     & $\phantom{}^{+7.6}_{-9.3}$  & $\phantom{}^{+63.0}_{-56.9}$\\
    \bottomrule
    \bottomrule
    \end{tabular}
    \caption{The uncertainties on the two different DM observables (see text), for two different annihilation channels for both a 100 GeV and 1000 GeV DM particle. The uncertainty on $\langle \sigma v \rangle$ is given \% since its absolute value only changes the normalization of the spectrum and is additionally model specific. All fits again only fit the spectrum at energies higher than 1 GeV.}
    \label{tab:resulting_uncertainties}
\end{table}
In table \ref{tab:resulting_uncertainties} we show the uncertainties for two fairly commonplace annihilation channels in DM studies; $b\bar{b}$ and $W^+W^-$, for two example masses, 100 GeV and 1000 GeV. Notably, the uncertainties for $b\bar{b}$ is fairly asymmetrical, this is mainly due to the fact that for $b\bar{b}$ the lower QCD uncertainties are larger than the upper uncertainties. Thus the specific annihilation channel is highly relevant for the resulting uncertainties on the DM observables, which can reach up to 14\% in certain scenarios.

\section{Conclusions}
\label{sec:conclusion}

In this work, we have shown for the first time that QCD uncertainties on antiproton spectra may have an important effect on the quality of DM fits. Therefore, these uncertainties are relevant for DM interpretations of the AMS--02 excess and should be used in future DM analyses\footnote{A. Jueid, J. Kip, R. Ruiz de Austri, P. Skands, in preparation.}. We analysed the effects of QCD uncertainties from both parton showers and hadronisation on the best-fit point and found that they can change the mass of DM by $\Delta M_\chi = \phantom{}^{+12.8}_{-4.5}~(\phantom{}^{+143.3}_{-65.5})$ GeV for $M_\chi = 100~(1000)$~GeV in the $b\bar{b}$ annihilation channel. The effects of the QCD uncertainties on the DM mass in the $W^+W^-$ annihilation channel are of similar size as in the $b\bar{b}$ channel. Moreover, we found that the thermally-averaged annihilation cross section is affected by $6.3\%$--$9.4\%$ depending on the DM mass and the annihilation channel.  The comparatively large size of the QCD uncertainties on antiproton spectra (relative, e.g., to those for photon spectra~\cite{Amoroso:2018qga}) are partly due to  tensions between different measurements of baryon spectra at LEP. Complementary detailed measurements of baryon spectra at the LHC could therefore potentially deliver additional constraints that would be indispensable not only for DM analyses in indirect-detection experiments but also for extrapolations of fragmentation functions needed for high-energy cosmic rays. The final states at LHC are of course far more complex than those at LEP, but relatively clean observables may still be constructed, e.g.~by using one of the recently developed quark/gluon jet discrimination observables for jets with very high transverse momenta \cite{Gallicchio:2011xq,Komiske:2016rsd,Metodiev:2018ftz,Gras:2017jty}.  To facilitate further full-fledged analyses and help the user to quickly estimate the effects of the QCD uncertainties on the fit results, we provide tabulated spectra with uncertainties and the code to determine DM observable uncertainties at \href{https://github.com/ajueid/qcd-dm.github.io.git}{github}. The tables can also be found in the latest releases of \textsc{DarkSusy}~6 \cite{Bringmann:2018lay},  \textsc{MicrOmegas}~5 \cite{Belanger:2018ccd} and \textsc{MadDM} \cite{Ambrogi:2018jqj}.

\acknowledgments

The work of AJ is supported in part by the Institute for Basic Science (IBS) under the project code, IBS-R018-D1. JK is supported by the NWO Physics Vrij Programme ``The Hidden Universe of Weakly Interacting Particles" with project number 680.92.18.03 (NWO Vrije Programma),
which is (partly) financed by the Dutch Research Council (NWO). R. RdA acknowledges the Ministerio de Ciencia e Innovación (PID2020-113644GB-I00). PS is funded by the Australian Research Council via Discovery Project DP170100708 — ``Emergent Phenomena in Quantum Chromodynamics''. This work was also supported in part by the European Union’s Horizon 2020 research and innovation programme under the Marie Sklodowska-Curie grant agreement No 722105 — MCnetITN3.

\bibliographystyle{JHEP}
\bibliography{bibliography.bib}

\end{document}